\documentclass{mem}
\usepackage{natbib}\usepackage{txfonts}\usepackage{balance}
\usepackage{graphicx}
\usepackage[a4paper]{hyperref}
\idline{00}{000}
\def \th {\thinspace}

\def\approxgt{\mathrel{\hbox{\rlap{\lower.55ex \hbox {$\sim$}} \kern-.3em \raise.4ex \hbox{$>$}}}}
\def\lesssim{\mathrel{\hbox{\rlap{\lower.55ex \hbox {$\sim$}} \kern-.3em \raise.4ex \hbox{$<$}}}}
\def\approxlt{\mathrel{\hbox{\rlap{\lower.55ex \hbox {$\sim$}} \kern-.3em \raise.4ex \hbox{$<$}}}}

\begin{document}
\title{Neutral versus ionized absorber as an explanation of the X-ray dippers}
\subtitle{}

\author{
M.\, Ba\l uci\'nska-Church\inst{1,2}
\and M. J.\,Church\inst{1,2} 
\and T.\, Dotani\inst{3,4}
\and T.\, Hirotsu\inst{3,4}
}

\offprints{M. Ba\l uci\'nska-Church}

\institute{School of Physics \& Astronomy,
University of Birmingham, Birmingham B15 2TT, UK
\and
Astronomical Observatory, Jagiellonian University,
ul. Orla 171, 30-244, Cracow, Poland
\and
Institute of Space \& Astronautical Science,
3-3-3 Yoshinodai, Sagamihara, Kanagawa 229-8510, Japan
\and
Department of Physics, Tokyo Institute of Technology, Ohokayama, Meguro,
Tokyo, 15-8551, Japan.
\email{mbc@star.sr.bham.ac.uk}}

\authorrunning{Ba\l uci\'nska-Church}
\titlerunning{Cool absorber dipping}

\abstract{We present results of the {\it Suzaku} observation of the dipping, periodically bursting
low mass X-ray binary XB\th 1323-619 in which we concentrate of the spectral evolution in dipping
in the energy range 0.8 - 70 keV. It is shown that spectral evolution in dipping is well-described
by absorption on the bulge in the outer accretion disk of two continuum components: emission of the 
neutron  star plus the dominant, extended Comptonized emission of the accretion disk corona (ADC). This
model is further supported by detection of a relatively small, energy-independent decrease of flux
above 20 keV due to Thomson scattering. It is shown that this is consistent with the electron scattering
expected of the bulge plasma. We address the recent proposal that the dip sources may be explained 
by an ionized absorber model giving a number of physical arguments against this model. In particular,
that model is inconsistent with the extended nature of the ADC for which the evidence is now overwhelming.
}
\maketitle{}

\section{Introduction}

The dipping sources form a class of about 10 low mass X-ray binaries (LMXB) exhibiting reductions 
in X-ray intensity at the orbital period; dipping may be shallow or may be 100\% deep in the
band 1 - 10 keV. Dipping in general leads to a hardening of the spectrum due to removal of
lower energy photons indicating that it is caused by photoelectric absorption. It has been generally
accepted since the 1980s that this takes place in the bulge in the outer accretion disk where the accretion flow from
the companion impacts (White \& Swank 1982; Walter et al. 1982). This plasma has a low ionization state
as situated far from the neutron star and so spectral modelling of dipping has employed neutral
absorber cross sections as these do not differ substantially from those of low ionization state ions.
Sources with more unusual spectral evolution have been investigated, such as 4U\th 1755-338
in which dipping is approximately energy independent (Church \& Ba\l uci\'nska-Church 1993) and X\th 1624-490 
(Church \& Ba\l uci\'nska-Church 1995) in which a spectral softening takes place. However, the spectral
evolution in these sources was shown to be still well-described by photoelectric absorption by cool
material providing there are two contiuum emission components in the sources, i.e. blackbody emission
from the neutron star, plus Comptonized emission of an accretion disk corona (ADC).
 
{\it ASCA} and {\it BeppoSAX} observations of the dipping sources XB\th 1916-053 led to proposal
of the ``progressive covering'' explanation of dipping (Church et al. 1997, 1998a,b). The spectral
evolution in dipping in this source and dip sources in general is relatively complex with parts
of the spectrum being absorbed while other parts at lower energies were not absorbed. This may
be explained in terms of the major Comptonized emission component being extended such that it
is gradually overlapped by the extended absorber so that in any stage of dipping
the part of the extended ADC emission covered by the bulge is absorbed producing the absorbed part
of the spectrum, while the uncovered part is not covered producing the unabsorbed part of the
spectrum. As dipping proceeds, this part becomes a smaller fraction of the total as observed
and as confirmed by spectral fitting.
Spectral modelling on this basis gave very good descriptions of dipping, for example,
in the {\it ASCA} observations of XB\th 1916-053 (Church et al. 1997) and of XBT\th 0748-676
(Church et al. 1998a) and of the {\it BeppoSAX} observations of XB\th 1916-053 (Church 
et al. 1998b). It has since been shown to apply to the dipping sources generally.

The extended nature of the Comptonizing ADC has since then been demonstrated in the dipping sources
using the technique of dip ingress timing. The time taken for the transition from non-dip
to 100\% deep dipping $\Delta t$ is proportional to the size of the extended emitter:
$r_{\rm ADC}$ = $\pi \, r_{\rm AD}\, \Delta t/ P$, where $r_{\rm ADC}$ is the radial extent of the
corona, $r_{\rm AD}$ is the radial extent of the accretion disk and $P$ is the orbital period.
Application of this technique to the dipping LMXB (Church \& Ba\l uci\'nska-Church 2004) gave
ADC radial sizes as shown in Fig. 1 between 20\th 000 and 700\th 000 km, depending on source luminosity
so that for a bright source ($L$ $\sim$ $1\times 10^{38}$ erg s$^{-1}$), the size is $\sim$500\th 000 km.
\begin{figure}[!ht]                                                   
\includegraphics[width=44mm,height=64mm,angle=270]{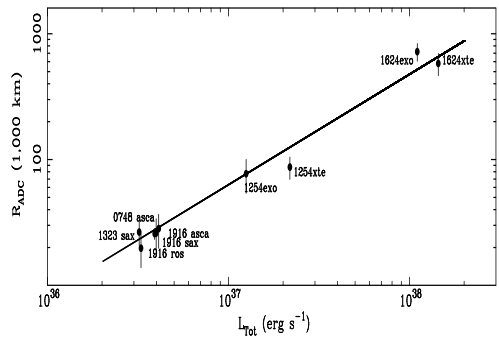} 
\caption{Measured values of the radial extent of the ADC in LMXB; from Church \& Ba\l uci\'nska-Church (2004).}
\end{figure}

Recently this has been confirmed by an independent technique based on {\it Chandra} data
by Schulz et al. (2009). Precise grating measurements of the spectrum of Cygnus\th X-2
reveal a wealth of emission lines of highly excited states such as the H-like ions of Ne, Mg, Si, 
S and Fe. The width of these lines
indicated Doppler shifts due to orbital motion in the accretion disk corona at radial positions
between 18\th 000 and 110\th 000 km in good agreement with the overall ADC size from dip ingress 
timing. The evidence for the extended corona is thus now overwhelming.

\subsection{The ionized absorber model}

In recent years, high resolution instruments on {\it XMM} and {\it Chandra} have revealed the presence 
of absorption lines of highly ionized species, for example in XBT\th 0748-676 (Jimenez-Garate et al.
(2003), and in X\th 1624-490 (Parmar et al. 2002). However, this proof of the existence of highly
ionized states was taken much further by Diaz-Trigo et al. (2006) and Boirin et al. (2005) who
proposed that \hbox{X-ray} 
\begin{figure*}[!ht]                                                   
\includegraphics[width=44mm,height=134mm,angle=270]{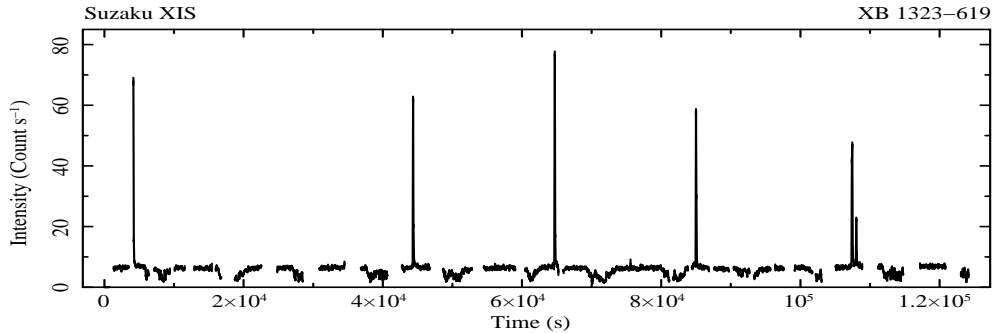} 
\caption{XIS lightcurve of the {\it Suzaku} observation of XB\th 1323-619 in the energy band
0.2 - 12 keV with 64 s binning.}
\end{figure*}
dipping in all LMXB dipping sources could be explained by an ionized absorber
model, i.e. showing that the spectra can be fitted with this model. Fitting the spectral evolution
in dipping with this model implies an absorbing medium of high and variable ionization state 
(log$\,\xi$ $\sim$3), which therefore cannot be in the bulge in the outer disk but must be located
much closer to the neutron star. In dipping log$\,\xi$ decreases from 3.5 to 3.0 while the column density
of ionized absorber increases $\sim$10 times, with very little change in the column density of neutral 
absorber.

However, the nature of the absorbing structure is not specified and the reason why the bulge in the outer
disk does not absorb is not addressed. The implications of this model, if correct, would be far-reaching 
for our understanding of LMXB. The proponents of the model argue that their fitting does not require
progressive covering, and therefore the model does not require there to be an extended corona.
Conversely, the evidence for the extended corona is now so strong that dipping cannot correctly be modelled
without progressive covering. 

In this work we present results of a recent observation of the dipping
source XB\th 1323-619 with {\it Suzaku}, viewing the results in the context of cool {\it versus}
ionized absorber.

\section{Observations and analysis}

We observed XB\th 1323-619 using {\it Suzaku} in 2007, Jan 9 - 10 for 34.1 hours, spanning
11.7 orbital cycles. Data from the XIS and HXD instruments were used. The XIS was operated
in the normal mode using the one quarter window mode viewing 1/4 of the CCD for pulse height
analysis with 2 s time resolution. Data were available for three detectors: XIS0, XIS1 and XIS3.
The data were filtered to remove hot and flickering pixels and on grade to discriminate
between X-ray and charged particle events. Standard screening was applied to remove periods of 
South Atlantic Anomaly, for elevation above the Earth's limb and for geomegnetic rigidity.
However, this
removed several X-ray bursts and so the screening criteria were relaxed, and it was found that 
this had negligible effect on the spectra. Full details are given in Ba\l uci\'nska-Church et al. (2009).

HXD data were also analyzed to allow broadband spectral fitting. In the PIN, the source 
was detected up to $\sim$50 keV. Because of this the GSO data with a band of 30 - 600 keV
were not used.

\vskip - 10mm
\section{Results}

The background-subtracted XIS lightcurve in the band 0.2 - 12 keV is shown in Fig. 1
with 64 s binning. Twelve dips can be seen and 5 bursts, one of which is double.
By removing the bursts completely and period searching, a best-fit orbital period 
could be obtained. This period was then refined by a technique of plotting the times
of dip minima against cycle number and fitting this, to give a best-fit period of
2.928$\pm$0.002 hr, in good agreement with the value previously obtained from
{\it BeppoSAX} of 2.938$\pm$0.020 hr (Ba\l uci\'nska-Church et al. 1999) and {\it Exosat}
2.932$\pm$0.005 hr (Parmar et al. 1989). In Fig. 3, we show the lightcurves in two energy bands:
1 - 4 keV and 4 - 12 keV folded on the best-fit period and the hardness ratio formed by
dividing these.

\begin{figure}                                                           
\includegraphics[width=64mm,height=64mm,angle=270]{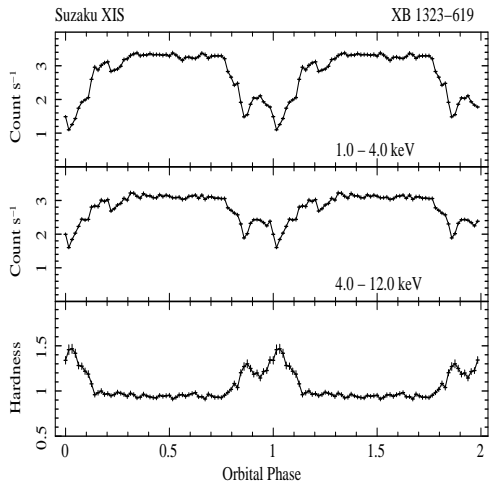} 
\caption{Background-subtracted XIS lightcurve of XB\th 1323-619 folded on the best-fit
orbital period of 2.928 hr obtained by dip analysis. Top panel: in the energy band 1.0- 4.0 keV;
middle panel in the band 4.0 - 12.0 keV; the lower panel shows the hardening in dipping 
in the ratio of the two energy bands.}
\end{figure}

\subsection{The non-dip spectrum}
 
Spectral analysis was carried out using a spectrum from which all traces of dipping
and bursting were removed. A PIN spectrum was extracted using the same selections
applied to the XIS data, and the XIS and PIN spectra were fitted simultaneously. Channels 
below 0.8 keV and above 12 keV in XIS were ignored, as were channels below 12 and above
70 keV in PIN. 

\begin{figure}                                                          
\includegraphics[width=64mm,height=64mm,angle=270]{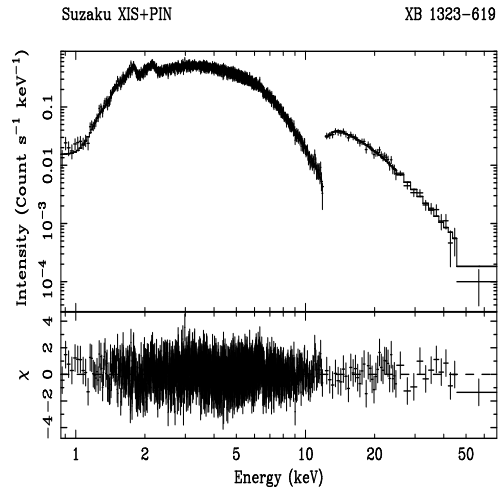} 
\caption{The non-dip spectrum: best-fit to the XIS and PIN detectors obtained from simultaneous fitting.}
\end{figure}

The spectrum was well-fitted by a model consisting of Galactic absorption, blackbody 
emission from the neutron star, Comptonized emission of the ADC plus a number of line features.
In Fig. 4 we show the best-fit to the XIS and PIN simultaneously and in Fig. 5
there is an expanded view of the structure seen between 5 and 10 keV. The best-fit was found for
a Galactic $N_{\rm H}$ of $3.2\times 10^{22}$ atom cm$^{-2}$, a blackbody temperature of 
1.35$\pm$0.36 keV, Comptonization described by a power law of photon index $\Gamma$ = 
1.67$^{+0.10}_{-0.03}$ and a high energy cut-off of 85$^{+77}_{-35}$ keV. An absorption line 
was found at 6.67 keV identified as Fe XXV; possible weak features at about 6.59, 6.74 and 
6.9 keV may also have been present. 

\begin{figure}                                                           
\includegraphics[width=64mm,height=64mm,angle=270]{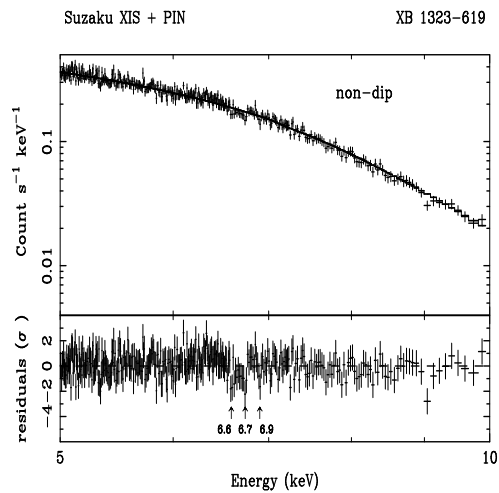} 
\caption{Expanded view of the non-dip XIS spectrum in the neighbourhood of the line features
detected (see text).}
\end{figure}

\subsection{The deep dip spectrum}

\begin{figure}                                                             
\includegraphics[width=64mm,height=64mm,angle=270]{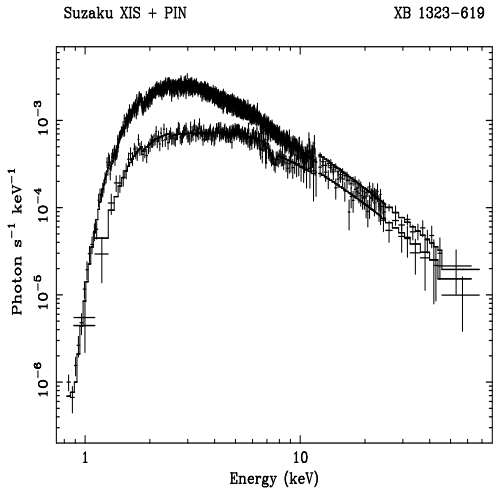} 
\caption{Comparison of the best-fits to the XIS plus PIN non-dip and deep dip spectra showing the 
the energy-independent reduction in flux at higher energies due to electron scattering.}
\end{figure}

A deep dip spectrum from both XIS and PIN was extracted by selecting for intensities less than 
1.2 count s$^{-1}$, in addition using only orbital phases between -0.16 and +0.09
at which deep dipping occurred as shown in the folded light curve (Fig. 3), and also removing all bursts. 
We tested the progressive covering model by applying this in the form
{\sc ag}$\ast$({\sc pcf}$\ast${\sc cut} + {\sc ag}$_{1}$$\ast${\sc bb)} i.e. applying the
covering factor {\sc pcf} to the cut-off power law modelling the extended emission of the ADC, while the
point-like blackbody {\sc bb} is subject to a column density {\sc ag}$_{1}$ in addition to the Galactic column density
{\sc ag}. All of the emission parameters (blackbody temperature etc) and {\sc ag} were
frozen at non-dip values as these cannot change in dipping. Acceptable fits were obtained but
the residuals clearly indicated absorption features and eventually, four absorption lines were added.
For a full discussion see Ba\l uci\'nska-Church et al. (2009). However, as seen in Fig. 6 where the
deep dip and non-dip spectra were compared it was clear that the deep dip spectrum at energies above 
20 keV was displaced vertically downwards, i.e. shifted in an energy-independent way indicating 
electron scattering. An additional factor had to be included for this which fitting showed to be
0.79$\pm$0.04. However, this may underestimate the error, which we checked by determining non-dip
and dip fluxes in several energy bands above 20 keV and a more reliable shift factor is 0.79$\pm$0.10.

We show below that this degree of electron scattering is consistent with that
expected in the plasma of the bulge. In the final fit of the deep dip spectrum, a very good fit
was obtained ($\chi^2$/d.o.f. = 140/263) in which the  extended ADC was subject to a covering fraction $f$ of 
0.63$\pm$0.02 with column density 22.2$\pm 0.8\times 10^{22}$ atom cm$^{-2}$ and the neutron star
blackbody was completely removed, being subject as a point source to the densest part of the
absorbing bulge in the outer disk. Absorption lines were seen at 6.52$\pm$0.10, 6.68$^{+0.18}_{-0.02}$,
6.94$\pm$0.14 and 7.6$\pm$0.2 keV (Ba\l uci\'nska-Church et al. 2009). 

\subsection{The vertical shift in dipping}

The expected reduction in X-ray intensity above 20 keV where photoelectric absorption is negligible
is $I/I_0$ =  $exp\, -N_{\rm H}\, \sigma_{\rm T}$ where $N_{\rm H}$ = $22\times 10^{22}$ atom cm$^{-2}$  is the 
measured additional column density for the Comptonized emission in dipping averaged over the
absorber, and $\sigma_{\rm T}$ is the Thomson cross section, giving $I/I_0$ = 0.86. However, only 63\%
of the emission is covered so that the overall factor $I/I_0$ = \hbox{0.86$\,$ x $\,$0.63 + 0.37} = 0.91. The 
measured factor of 0.79$\pm$0.10 is marginally consistent with this, although the shift is somewhat 
greater than expected from the above calculation. There will be some error in using neutral absorber 
photoelectric cross sections for the bulge. From the measured luminosity and column density of the ADC
in dipping, we find an ionization parameter $\xi$ $\sim$10 - 50 for the bulge, 
thus the ionization state in oxygen, for example, may be OII - OVI and the cross
sections may be up to a factor of two smaller than in neutral material. Thus the above calculation 
would require a higher value of $N_{\rm H}$ and the value of $I/I_0$ expected would be closer to the
measured value. It thus appears that the degree of electron scattering detected is consistent with 
scattering within the plasma of the bulge in the outer accretion disk.

\section{Discussion}

Our analysis shows that dipping in XB\th 1323-619 is well-described as absorption in
cool material supporting the standard view that dips are generated in the bulge in the
outer accretion disk. Moreover, we are able to detect the expected effect of electron scattering
in this low ionization state plasma as a vertical, energy-independent shift in the deep dip
spectrum at energies above 20 keV where photoelectric absorption will not contribute. 

\subsection{Arguents against the ionized absorber model}

As the standard view of the dip sources has been accepted generally
since the mid-1980s, it may be helpful to summarise arguments against the ionized
absorber model proposed as an alternative to the standard model.
Firstly, this model requires a time-varying, and orbital linked, variation of ionization state
at some position in the inner disk which has not been explained within the model.
Secondly, the azimuthal structure responsible for dipping over the observed ranges
of orbital phase is not explained. In addition to this, the reason why the bulge in the
outer disk is {\it not} involved in dipping is not explained. It is clear that
the bulge cannot be in an extremely high ionization state and therefore photolelectric
absorption must take place. This could be only avoided if the inclination angle of the dipping
sources was not high contrary to strong evidence such as the eclipses seen in some sources.

Set against these difficulties must be the evidence that dipping can be explained by
absorption in the bulge 
i.e. by an extended absorber moving across
an extended Comptonizing ADC (Church et al. 1997, 1998a,b; Ba\l uci\'nska-Church et al. 1999)
also removing the point-source neutron star emission when 
this is overlapped. In particular, the spectral anlysis presented here of dipping in XB\th 1323-619
strongly supports this view in detail, so that the degree of electron scattering observed is
also consistent with this explanation.

\section{Conclusions}

There are strong objections to the ionized absorber model as detailed above.
The standard explanation as absorption in the bulge does not have these objections
and is supported by the present work. The present work supports the
extended nature of the Comptonizing ADC for which there is now overwhelming evidence.

\begin{acknowledgements}

This work was supported in part by the Polish Ministry of Science and Higher Education 
grant 3946/B/H03/2008/34

\end{acknowledgements}

\bibliographystyle{aa}

\end{document}